\documentclass[a4paper]{jpconf}
\usepackage{graphicx}
\begin{document}

\newcommand{\Teff}{\ensuremath{T_{\mathrm{eff}}}}
\newcommand{\logg}{\ensuremath{\log g}}
\newcommand{\Fe}[5]{\mbox{$#1\,^#2{\rm #3}^{{\rm #4}}_{\rm #5}$}}
\newcommand{\Te}{T_{\rm e}}             
\newcommand{\mA}{{\rm m\AA}}       
\newcommand{\Elow}{E_{\rm low}}     
\newcommand{\EW}{W_{\lambda}}
\newcommand{\opd}{\log \tau_{\rm 5000}}

\title{NLTE effects on Fe I/II in the atmospheres of FGK stars and application to
abundance analysis of their spectra}
\author{Maria Bergemann$^1$, Karin Lind$^1$, Remo Collet$^{1}$ and Martin
Asplund$^1$}
\address{$^1$ Max-Planck Institute for Astrophysics, Karl-Schwarzschild Str. 1,
85741, Garching, Germany}
\ead{mbergema@mpa-garching.mpg.de}

\begin{abstract}
We describe the first results from our project aimed at large-scale calculations of NLTE abundance corrections for important astrophysical atoms and ions. In this paper, the focus is on Fe which is a proxy of stellar metallicity and is commonly used to derive effective temperature and gravity. We present a small grid of NLTE abundance corrections for Fe I lines and discuss how NLTE effects influence determination of effective temperature, surface gravity, and metallicity for late-type stars.
\end{abstract}

\section{Introduction}
Iron is a key element in stellar astrophysics. The very complex atomic structure of its lowest ionization stages, Fe I and Fe II, gives rise to a wealth of spectral lines all across the spectrum of a typical late-type star. This atomic property, coupled to a relatively large abundance makes Fe a reference for spectroscopic estimates of stellar parameters using the method of excitation-ionization balance.

The classical implementation of this method in spectrum analysis codes involves three assumptions: local thermodynamic equilibrium (LTE), hydrostatic equilibrium, and 1D geometry. These approximations strongly reduce the complexity of the problem, thus  permitting analysis of very large stellar samples in short timescales. Yet, in the conditions when the breakdown of 1D static LTE models occurs the inferred stellar parameters suffer from large systematic biases. To assess the latter, more physically realistic modeling is necessary.

Studies of NLTE effects on the Fe I/Fe II level populations for FGK stars trace back to Athay \& Lites (1972). Since then, vast amount of work has been performed in this field demonstrating that NLTE effects in the excitation-ionization balance of Fe I/Fe II are significant and can not be ignored even in the analysis of solar-type stars (Mashonkina et al. 2011 and references therein). Yet, despite major efforts aimed at understanding how non-equilibrium thermodynamics affects the line formation of Fe, there has never been an attempt to accurately quantify these deviations in a systematic manner across a wide range of stellar parameters, and to apply them to a large stellar sample. 

Here we present a new NLTE model atom of Fe I/Fe II to be applied in large-scale calculations of NLTE abundance corrections for late-type stars. We discuss the NLTE effects influencing atomic level populations at the typical conditions in their atmospheres and provide a small grid of NLTE corrections for the Fe I lines. The model has been tested on a number of well-studied stars with independently-determined parameters, including metal-poor giants and turnoff stars. These tests performed with classical 1D hydro-static model atmospheres and averages of 3D hydrodynamical simulations of stellar convection will be presented in Bergemann et al. (in preparation). A complete grid of NLTE abundance corrections computed with \textsc{multi2.3} statistical equilibrium code will be presented in Lind et al. (in preparation).

\section{Methods}{\label{sec:methods}}

\subsection{Model atmospheres and codes}

The calculations presented here were performed with 1D LTE plane-parallel \textsc{mafags-odf} models (Fuhrmann et al. 1997, Grupp 2004). In these models, convective energy transport is accounted for using the mixing-length theory of B\"ohm-Vitense (1958) with the mixing length parameter $\alpha_{\rm mlt}$ set to $0.5$. Line opacity is represented by all elements up to Ni, and various diatomic molecules (H$_2$, CH, CO, TiO, etc). This accounts for nearly 20 million atomic and molecular lines. The reference solar abundances were compiled from various literature sources, giving preference to the determinations by the Munich group. The model atmosphere provide partition functions and partial pressures of all relevant atoms and molecules, which are then used in detailed line formation calculations.

NLTE statistical equilibrium was computed with a revised version of the \textsc{detail} code (Butler \& Giddings 1985), which solves multi-level NLTE radiative transfer with a given static 1D model atmosphere. The last version of the code is based on the method of Accelerated Lambda Iteration (ALI). LTE and NLTE line formation calculations with departure coefficients from \textsc{detail} were performed with the revised version of the spectrum synthesis code \textsc{SIU} (Reetz 1999). The major update in the code relates to the automated computation of NLTE abundance corrections\footnote{NLTE abundance correction is defined to be the difference in abundance required to match NLTE and LTE line profiles or equivalent widths}, which generally proceeds by constructing the grids of LTE and NLTE line equivalent widths for a range of stellar parameters and interpolating in the corresponding curves-of-growth for the input observed equivalent widths or abundances.

\subsection{Model atom}
The NLTE model atom of Fe was constructed using all available atomic data from the Kurucz\footnote{http://kurucz.harvard.edu/atoms.html} database, which also includes laboratory data from the NIST compilation\footnote{http://www.nist.gov/pml/data/asd.cfm}. All predicted energy levels of Fe I with the same parity above $\Elow \geq 5.1$ eV were grouped into super-levels. Transitions between the components of the super-levels were also grouped. The total transition probability of a super-line is a weighted average of $\log gf$'s of individual transitions. Thus, in the final atomic model, the
number of energy levels is $296$ for Fe I and $112$ for Fe II, with uppermost excited levels located at $0.03$ eV and $2.72$ eV below the respective ionization limits, $7.9$ eV and $16.19$ eV. The model is closed by the Fe III ground state. The total number of radiatively-allowed transitions is $16\,207$
($13\,888$ Fe I and $2\,316$ Fe II).
\begin{figure}[h]
\hbox{
\resizebox{0.5\columnwidth}{!}{\rotatebox{90}
{\includegraphics[scale=1]{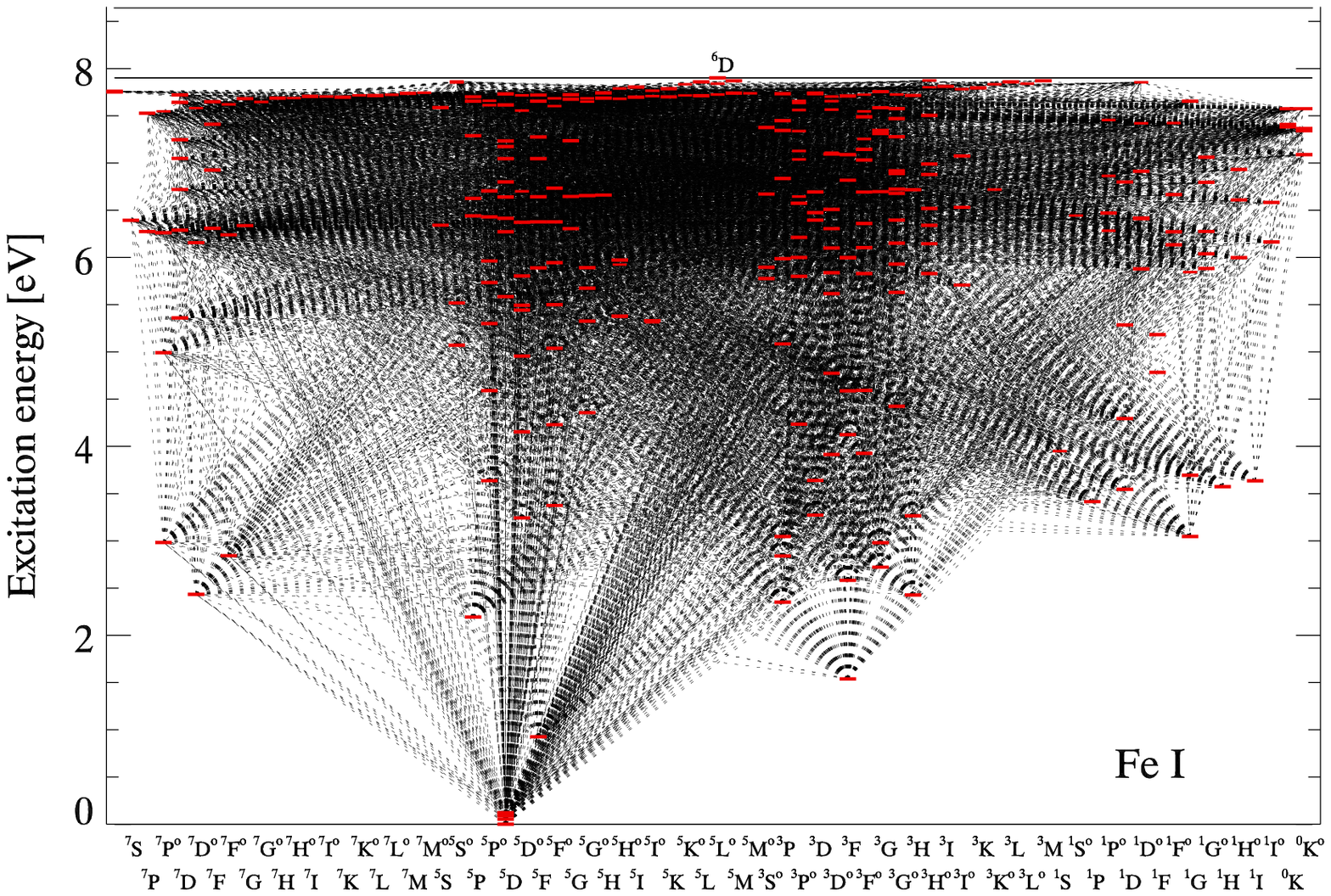}}}\hfill
\includegraphics[scale=0.4]{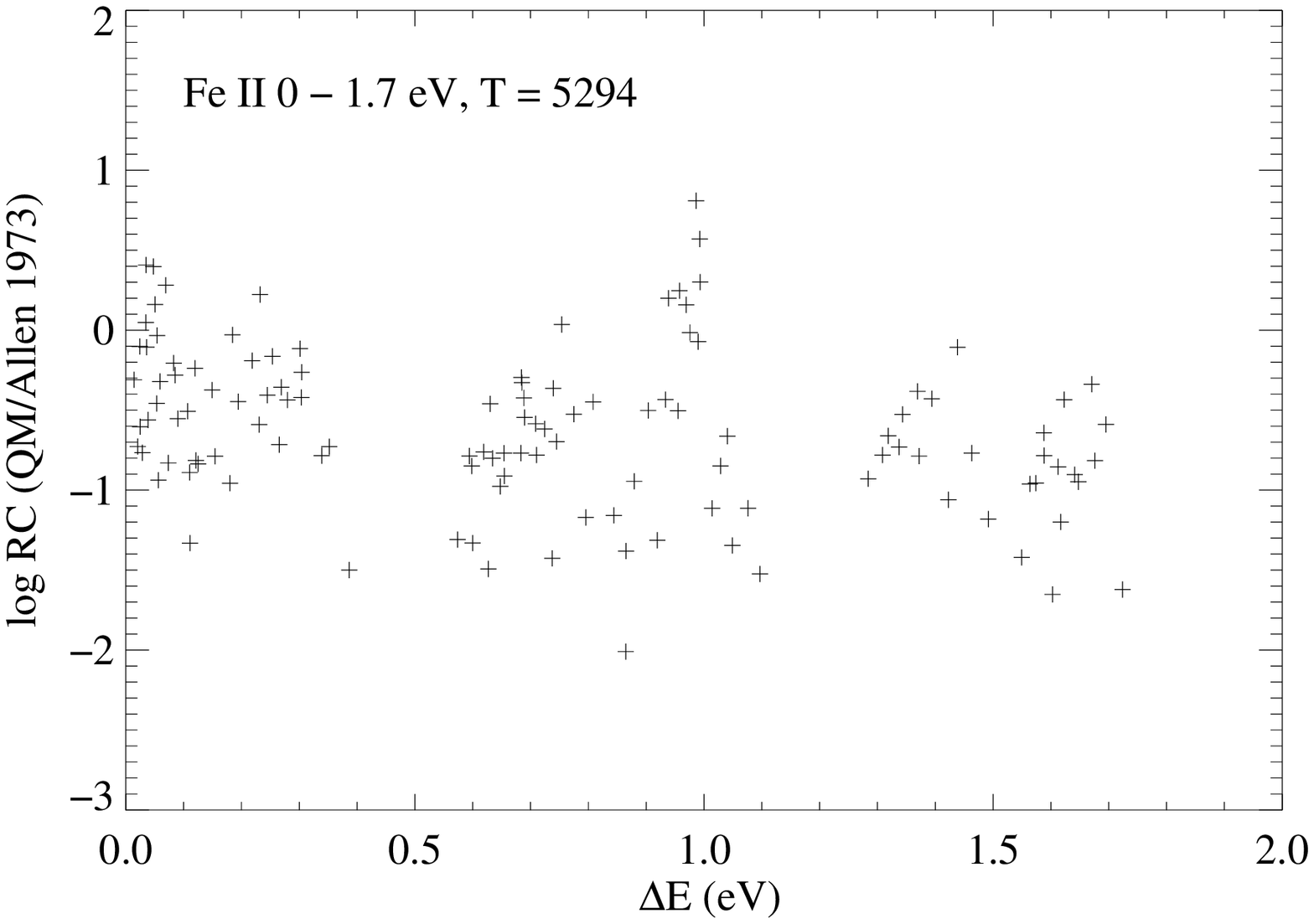}}
\caption{\label{atom} Left panel: Grotrian diagram of the Fe I atom. Right panel: comparison of quantum-mechanical rate coefficients for e$^-$-induced forbidden transitions between the lowest energy levels of Fe II with the classical Allen's (1973) recipe.}
\end{figure}

Photoionization cross-sections for $136$ states of Fe I were taken from Bautista (1997) and the hydrogenic approximation was adopted for all other levels. The rates of transitions induced by inelastic collisions with free electrons (e$^-$) and H I atoms were computed using different recipes. For the states coupled by permitted b-b and b-f transitions, we used the formula of van Regemorter (1962), respectively, Seaton (1962) for e$^-$.
To derive the rates of transitions due to inelastic collisions with H I atoms, we used the DrawinÕs formula (Drawin 1969) in the version of Steenbock \& Holweger (1984). Also, the states are connected by forbidden transitions induced by inelastic collisions with $e^-$ and H I atoms. These are computed using the formulae of Allen (1973) and Takeda (1994). Quantum-mechanical calculations exist only for e$^-$-induced forbidden transitions between the 16 lowest Fe II energy levels (Ramsbottom et al. 2007). These data are compared with the Allen's (1973) recipe in Fig. \ref{atom}.

The influence of e$^-$ and H I collisions on the statistical equilibrium of Fe was carefully investigated by performing test calculations with various scaling factors to the above-mentioned formulae. The NLTE synthetic profiles were furthermore compared with the observed stellar spectra to check how well the test model atoms satisfy the constraint of ionization-excitation balance of Fe I/Fe II under restriction of different stellar parameters. The final choice of scaling factors will be discussed in Bergemann et al. (in preparation).

\section{Statistical equilibrium of Fe}

\subsection{NLTE effects}
\begin{figure} 
\hbox{
\resizebox{0.55\columnwidth}{!}{\includegraphics[scale=1.2]
{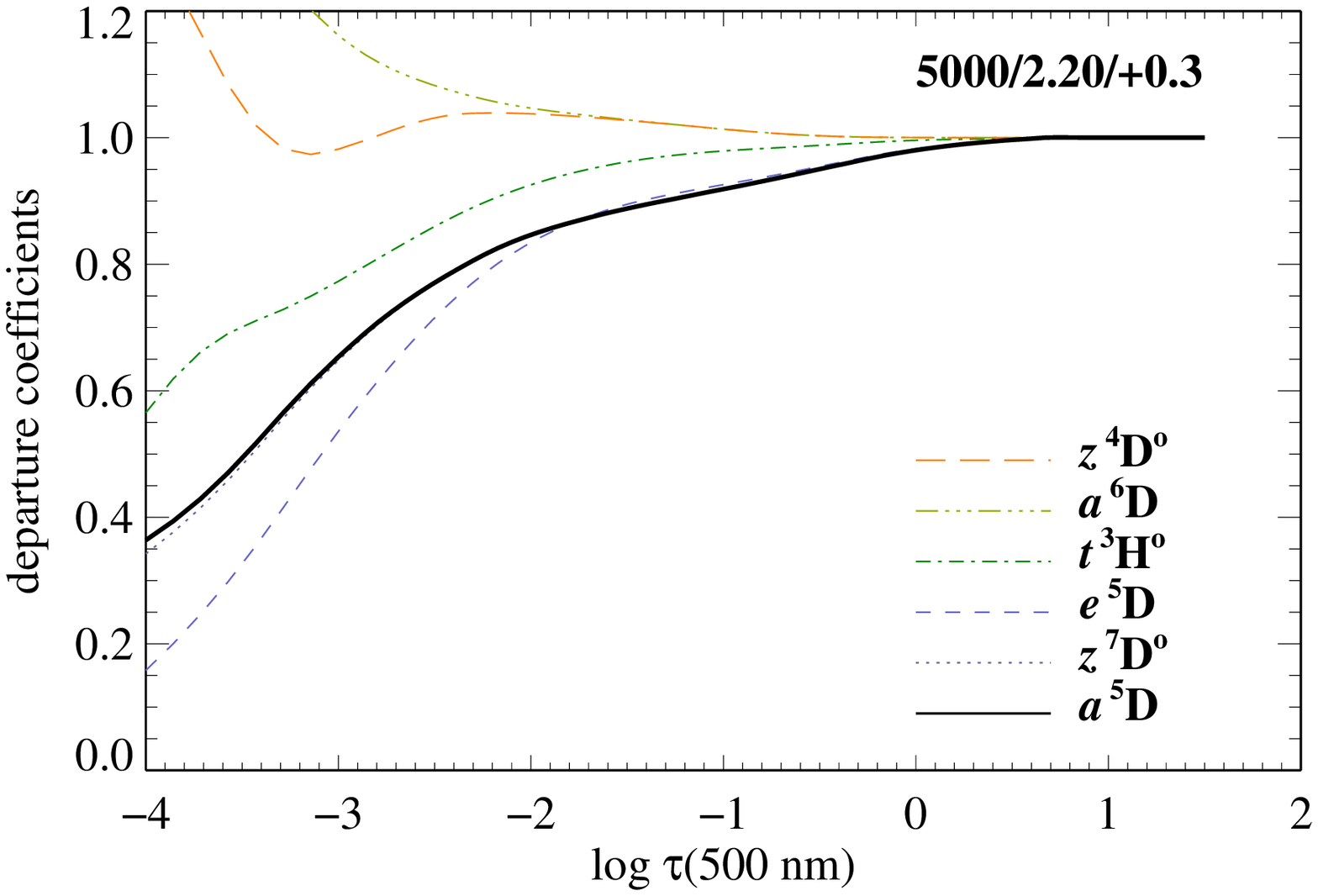}}\hfill
\resizebox{0.55\columnwidth}{!}{\includegraphics[scale=1.2]
{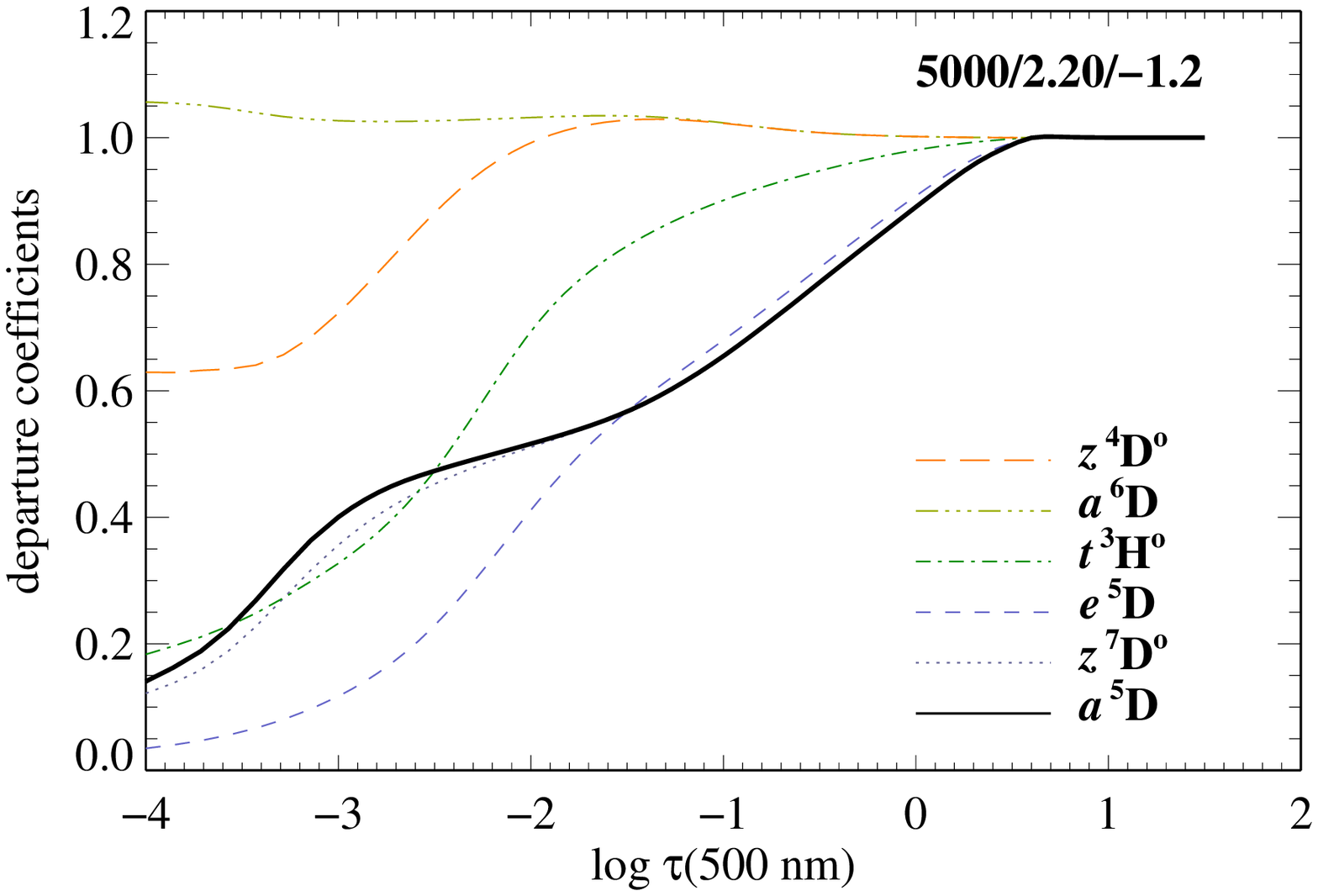}}}
\hbox{
\resizebox{0.55\columnwidth}{!}{\includegraphics[scale=1.2]
{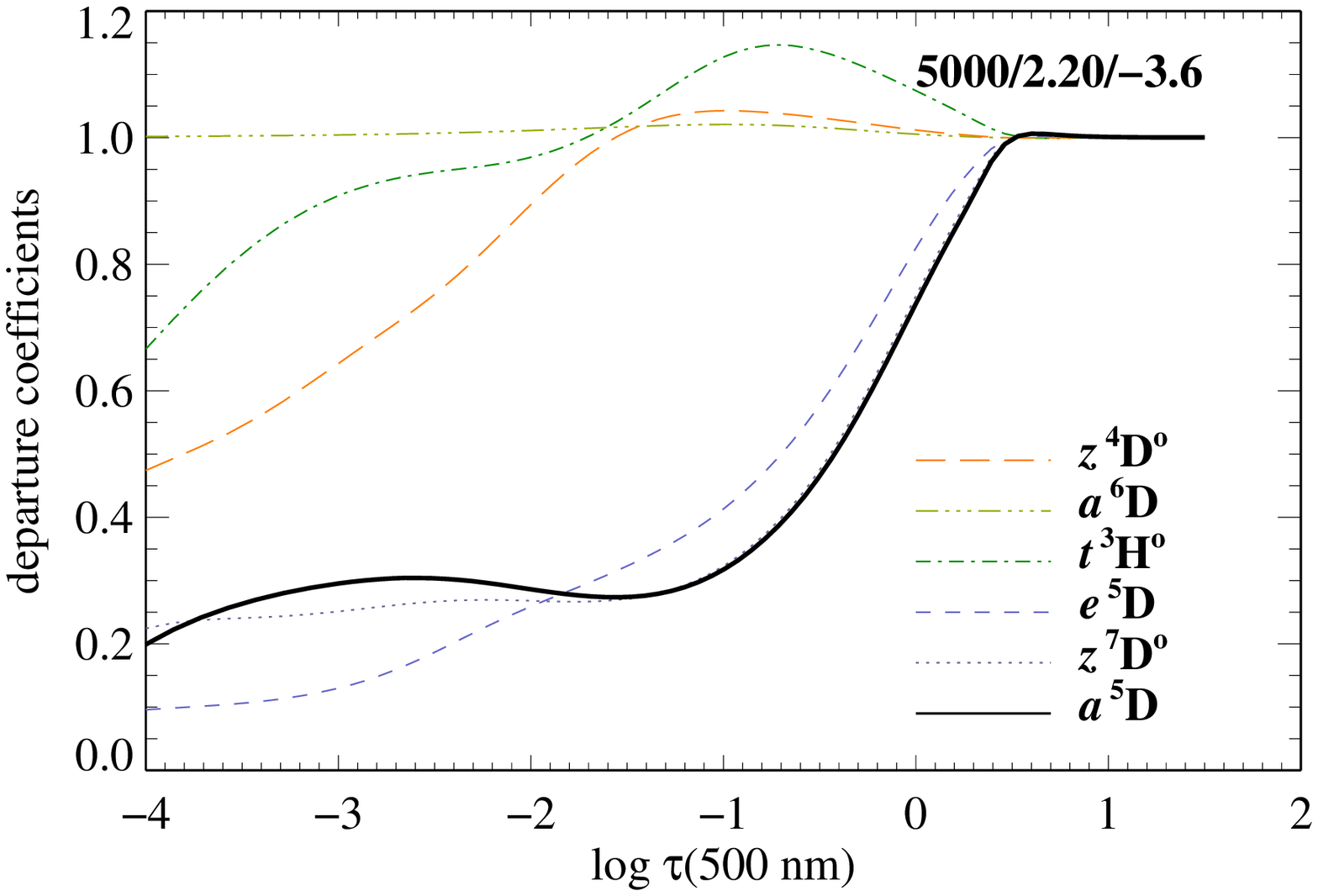}}\hfill
\resizebox{0.55\columnwidth}{!}{\includegraphics[scale=1.2]
{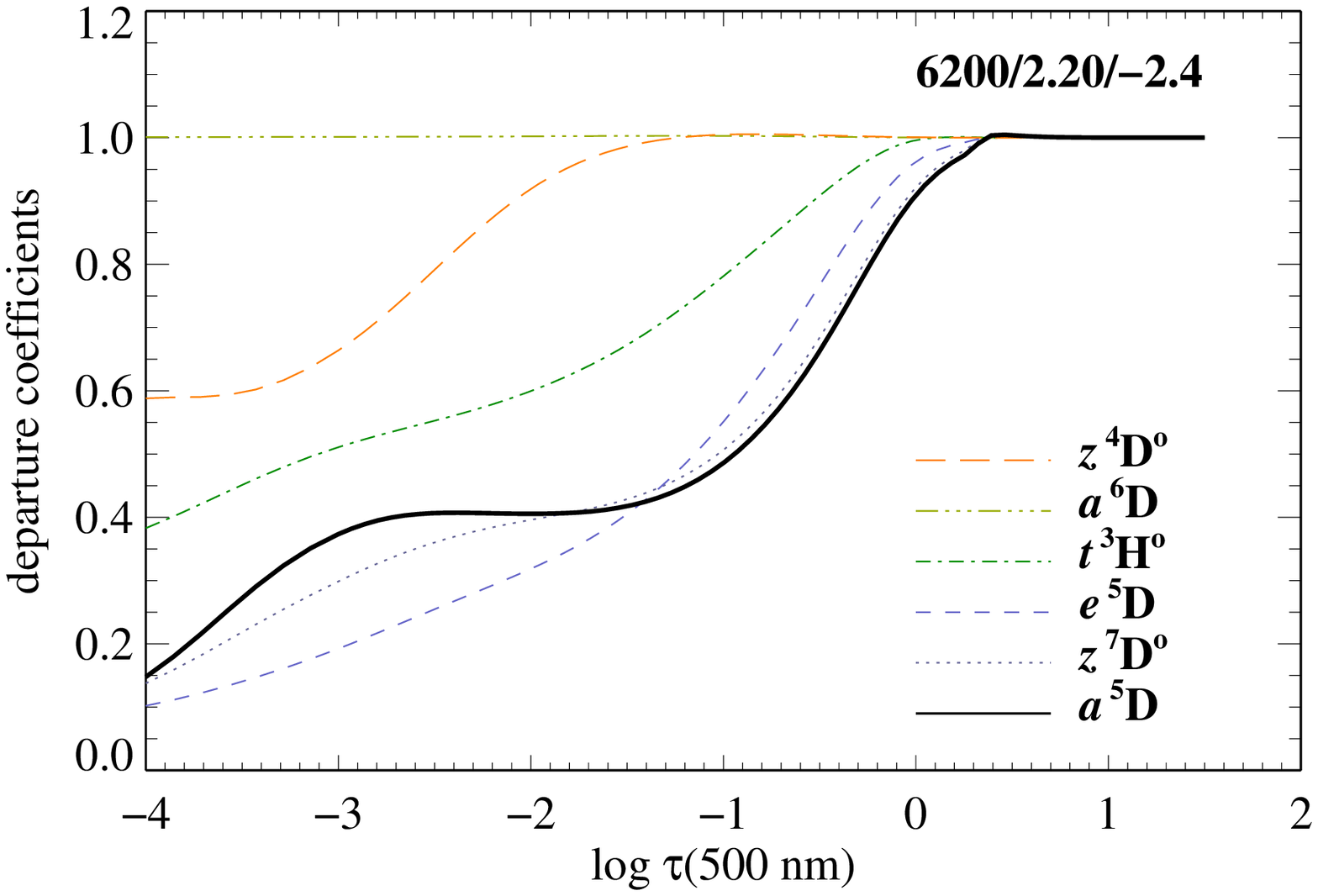}}}
\hbox{
\resizebox{0.55\columnwidth}{!}{\includegraphics[scale=1.2]
{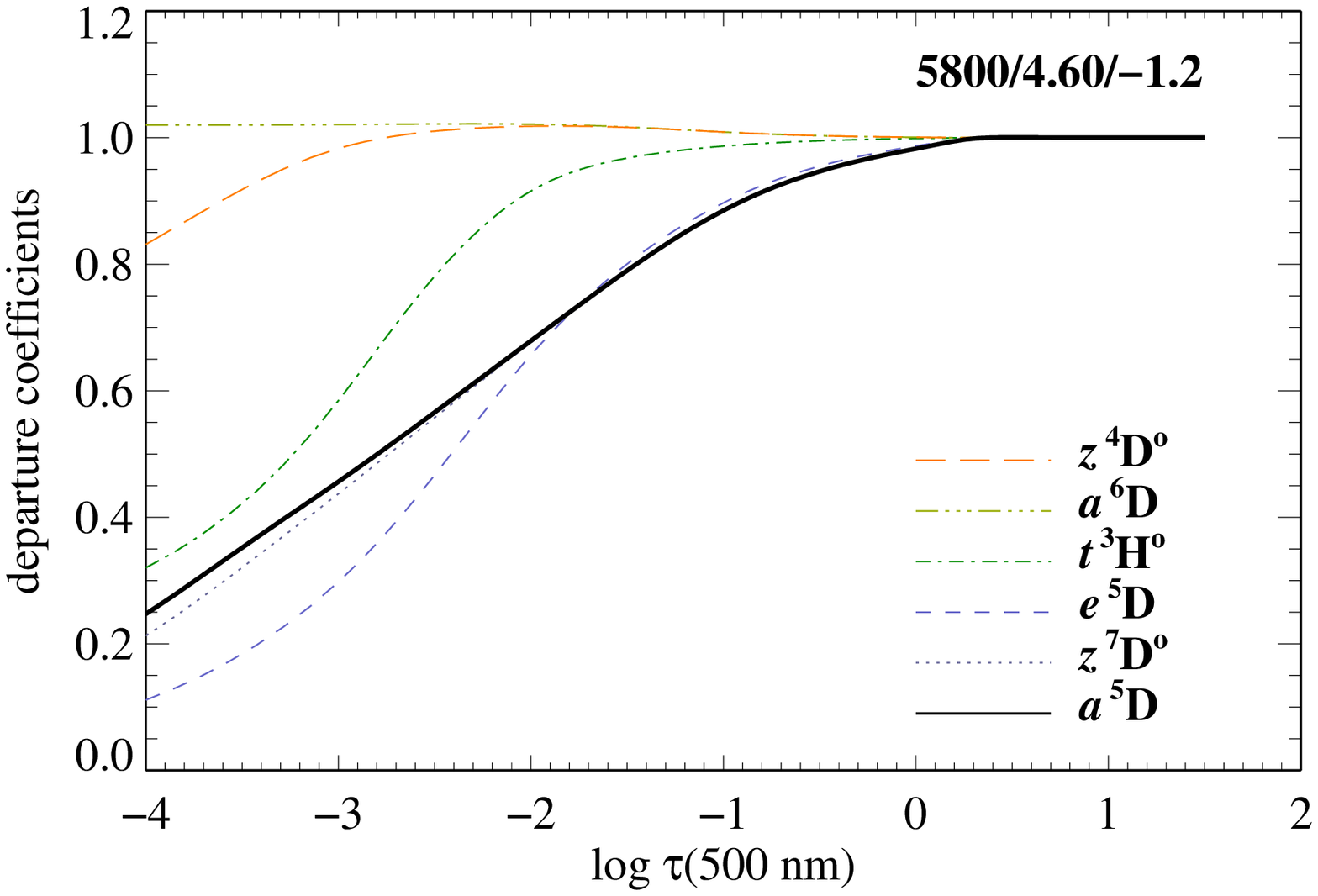}}\hfill
\resizebox{0.55\columnwidth}{!}{\includegraphics[scale=1.2]
{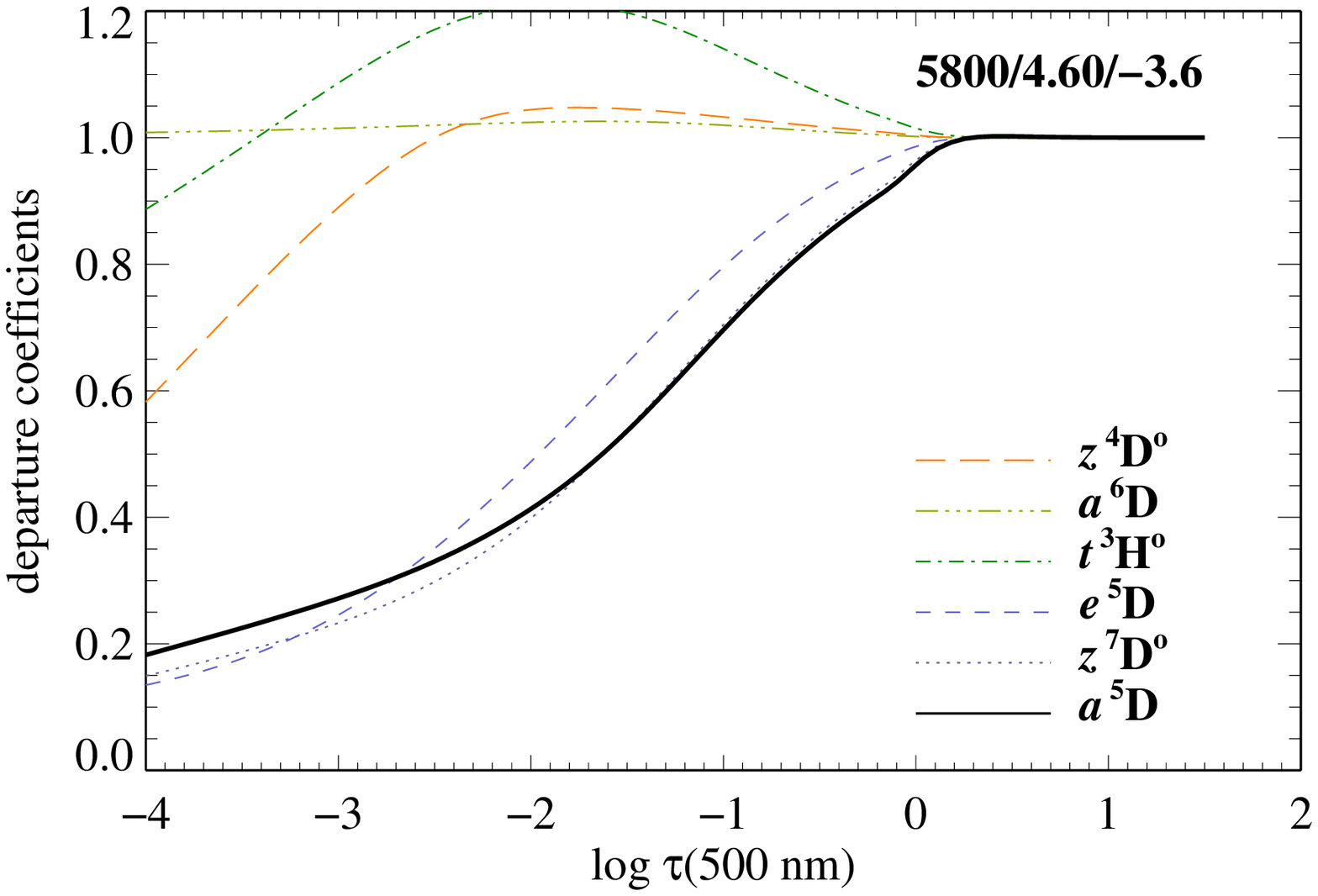}}} 
\caption{Departure coefficients of selected Fe I and Fe II levels for a number of model atmospheres from the grid; stellar parameters are indicated in each figure.} 
\label{dep} 
\end{figure}
Fig. \ref{dep} shows level departure coefficients, $b_i$,\footnote{Departure coefficient is defined 
to be the ratio of NLTE to LTE atomic level populations, $b_i = n_i^{\rm NLTE}/n_i^{\rm LTE}$} as a function of continuum optical depth $\opd$ for a number of model atmospheres computed with different stellar parameters. The latter are representative of stars typically used in Galactic chemical evolution studies. Only selected energy levels of Fe I and Fe II typical for their depth-dependence are included in the plot: \Fe{a}{5}{D}{}{} (ground state of Fe I), \Fe{z}{7}{D}{\circ}{} ($2.4$ eV), \Fe{e}{5}{D}{}{} ($5.4$ eV), \Fe{t}{3}{H}{}{\circ}, \Fe{a}{6}{D}{}{} (ground state of Fe II), and \Fe{z}{4}{D}{\circ}{} ($5.5$ eV). All these levels give rise to radiatively-permitted transitions, which are used in the detailed abundance analysis of the Sun and a number of metal-type stars. 

The common result for all the models shown in Fig. \ref{dep} is that in the optically thin atmospheric layers the majority of Fe I levels are underpopulated compared to LTE, $b_i < 1$. The Fe II number densities remain close to LTE values, $b_i \approx 1$, and a minor overpopulation of the Fe II ground state develops only close to the outer atmospheric boundary.	

Deviations from LTE in the distribution of atomic level populations arise because mean radiation field, $J_\nu$, at different depths and frequencies is not equal to the Planck function, $B_\nu [T_{\rm e}(\tau)]$, which is defined by the local temperature $T_{\rm e}(\tau)$ at each depth. For Fe I, excess of the mean intensity over Planck function in the UV continua leads to over-ionization, which sets in as soon as the optical depth in the photoionization continua of low-excitation Fe I levels falls below unity, i.e. $\opd \sim 0$. In the outer layers, ionization balance is dominated by the Fe I levels with excitation energies at $4 -5$ eV. In the infra-red continuum, $J_\nu < B_\nu$ leads to over-recombination, which is very efficient for our atomic model with only $0.03$ eV energy gap of the upper Fe I levels from the Fe II ground state.

Due to an extremely tight radiative and collisional coupling of atomic energy levels, excitation balance of Fe I is also affected by radiative imbalances in line transitions. These include radiative pumping by super-thermal UV radiation field of non-local origin at the line frequencies, as well as photon suction driven by photon losses in large-probability transitions between highly-excited levels. These processes leave a characteristic imprint on the behavior of $b_i$-factors in the outer atmospheric layers. In Fig. \ref{dep}, this is seen as the onset of deviations from the relative thermal equilibrium between the Fe I levels, which occurs, depending on the model metallicity, at $\opd \sim -1 ... -2$. The NLTE effects on the line formation are different for the strong and weak Fe I lines. The former are mostly shaped by deviations of their source function from the Planck function, which makes a clear effect in the line cores. Weak lines are pre-dominantly controlled by their opacity, thus the major change is due to the shift of their optical depth scale in NLTE.

\subsection{NLTE abundance corrections}

\begin{figure}[h]
\hbox{
\includegraphics[scale=0.4]{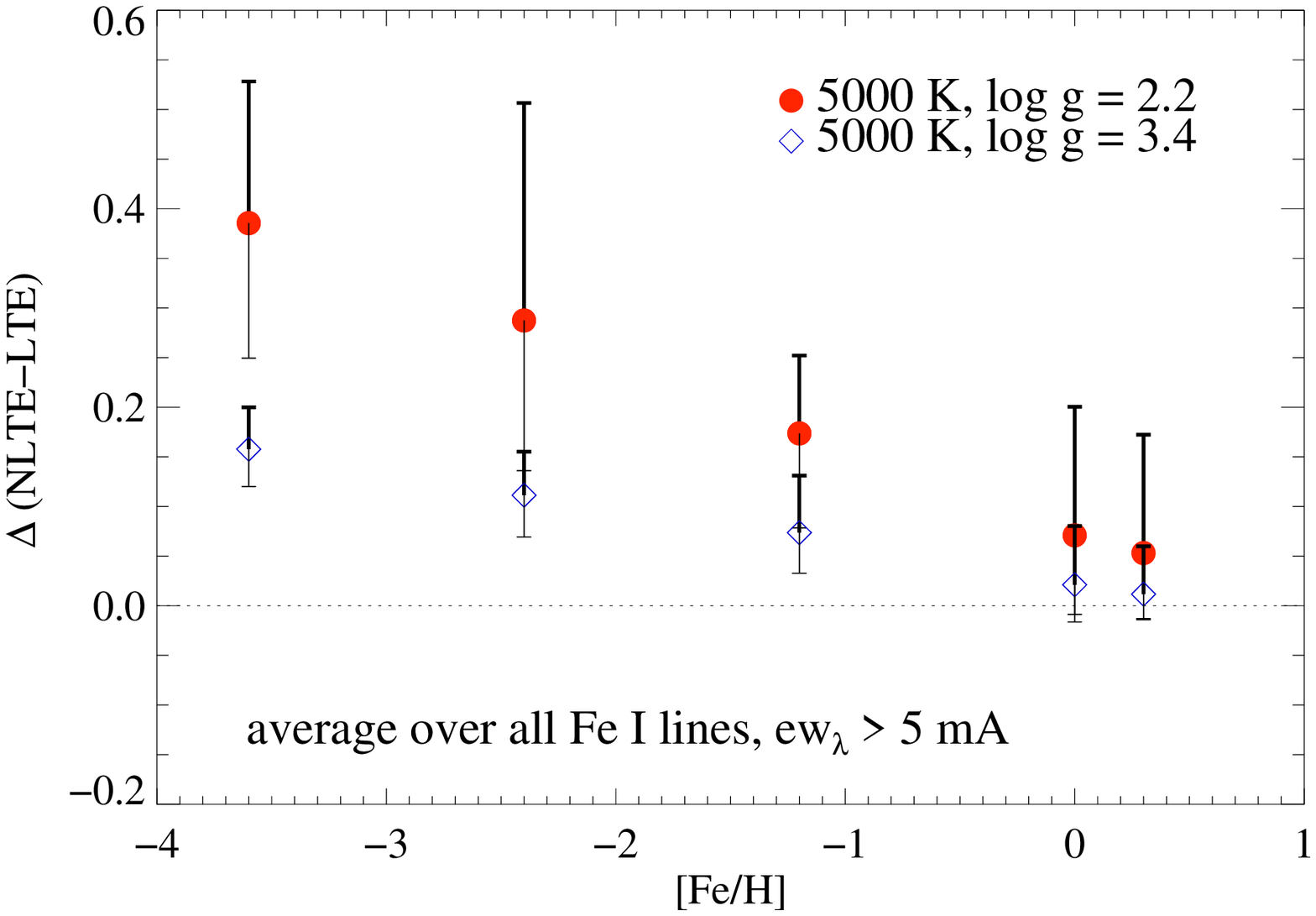}\hfill
\includegraphics[scale=0.4]{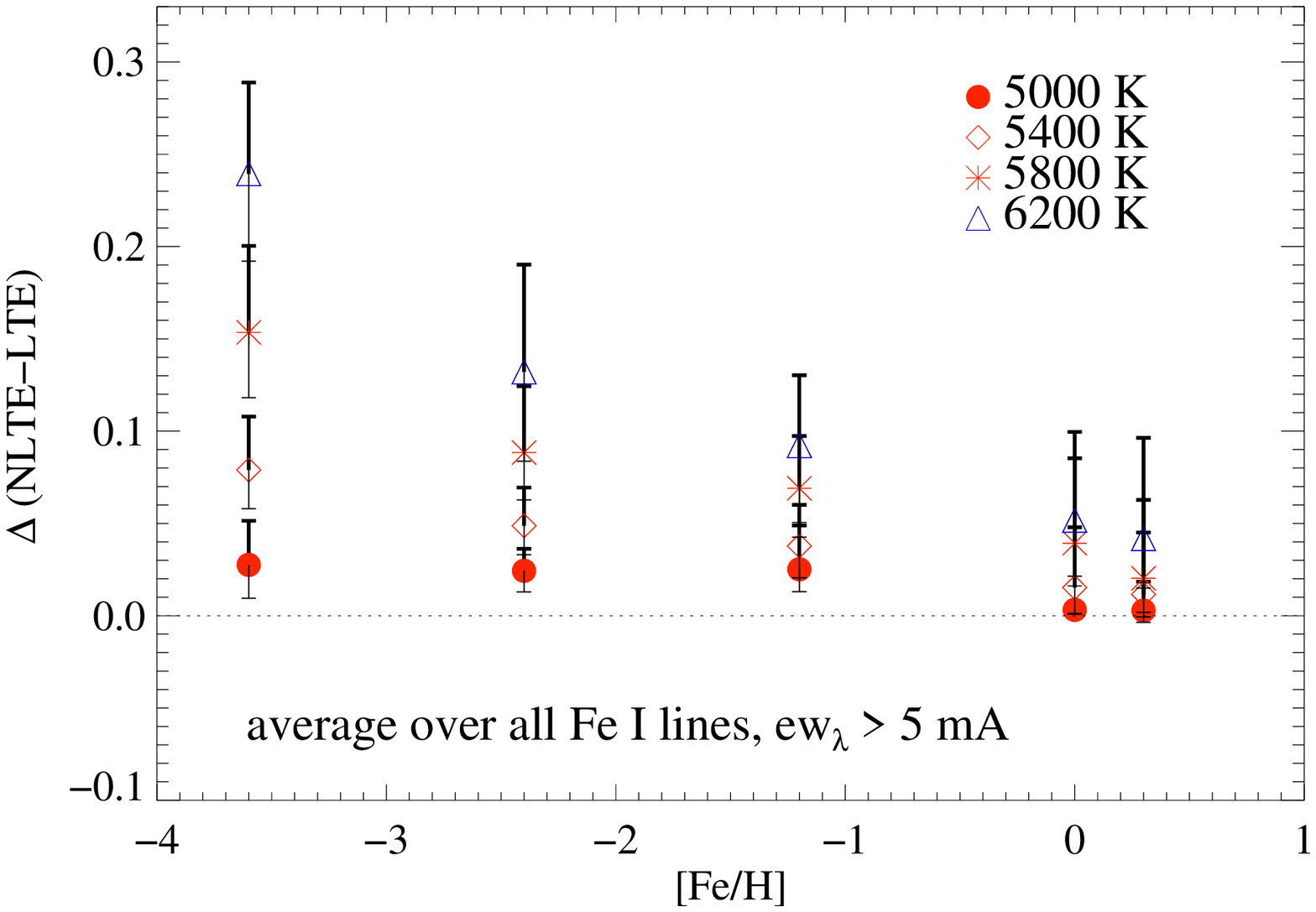}
}
\caption{\label{nlte}NLTE abundance corrections for different stellar parameters. Right panel: all model  atmospheres adopt $\logg = 4.6$.}
\end{figure}

We computed NLTE abundance corrections, hereafter $\Delta_{\rm NLTE}$, to Fe I lines for a small grid of model atmospheres covering metallicity range $-3.6 \leq $ [Fe/H] $ \leq +0.3$. The results are presented in Fig. \ref{nlte} for the parameters representative of giants (left panel) and dwarfs (right panel). Our sample of Fe I lines includes transitions between $0$ to $5$ eV levels, and the line equivalent widths range from $5$ \mA\ to few \AA\ depending on stellar parameters.
In the plot, each symbol corresponds to a mean $\Delta_{\rm NLTE}$ obtained by averaging individual NLTE corrections for all Fe I lines in the sample, and the errors bars correspond to the line-to-line variations in $\Delta$ for each model atmosphere. The former quantity roughly demonstrates how much the Fe I/Fe II ionization balance, and, thus, determination of surface gravity, is affected by NLTE, whereas the latter helps to assess the influence of NLTE on the excitation balance and determination of effective temperature. 

NLTE corrections increase with decreasing [Fe/H] and $\log g$, thus ionization balance achieved assuming LTE leads to progressively underestimated gravities and metallicities. In particular, at $\Teff = 5000$ K (Fig. \ref{nlte}, left panel), the mean NLTE corrections are twice as large in the $\log g = 2.2$ model compared to the $\log g = 3.6$ model, reaching $+0.4$ dex at [Fe/H] $\sim -3.5$. It is also interesting that NLTE corrections for dwarf models critically depend on $\Teff$. Fig. \ref{nlte} (right panel) shows that for the coolest models, $\Teff \sim 5000$ K, one may safely adopt LTE approximation down to the lowest metallicities. However, LTE leads to severe errors in gravity estimates for warm turn-off stars, $T > 6000$ K.

\begin{figure}[h]
\hbox{
\includegraphics[scale=0.4]{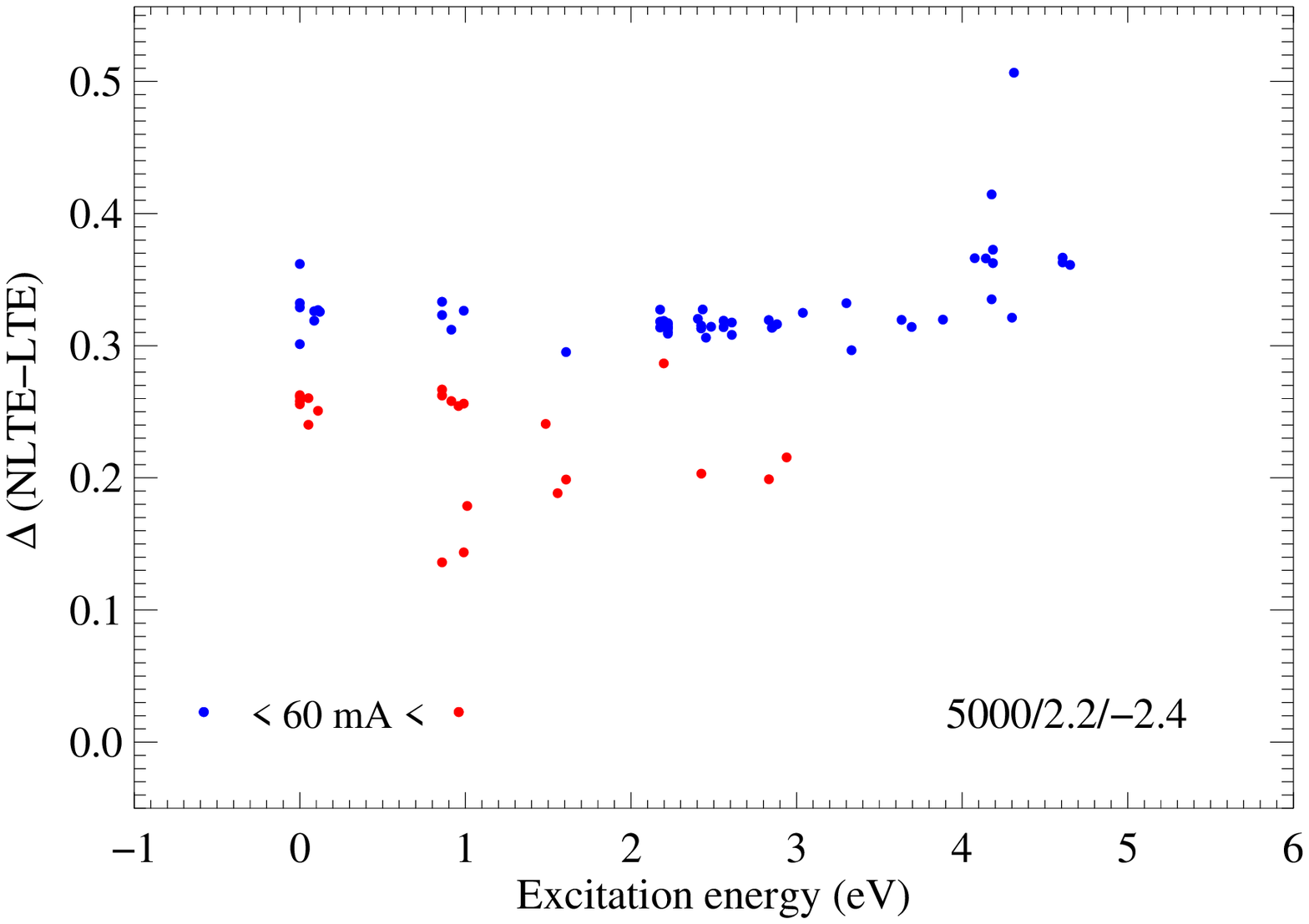}\hfill
\includegraphics[scale=0.4]{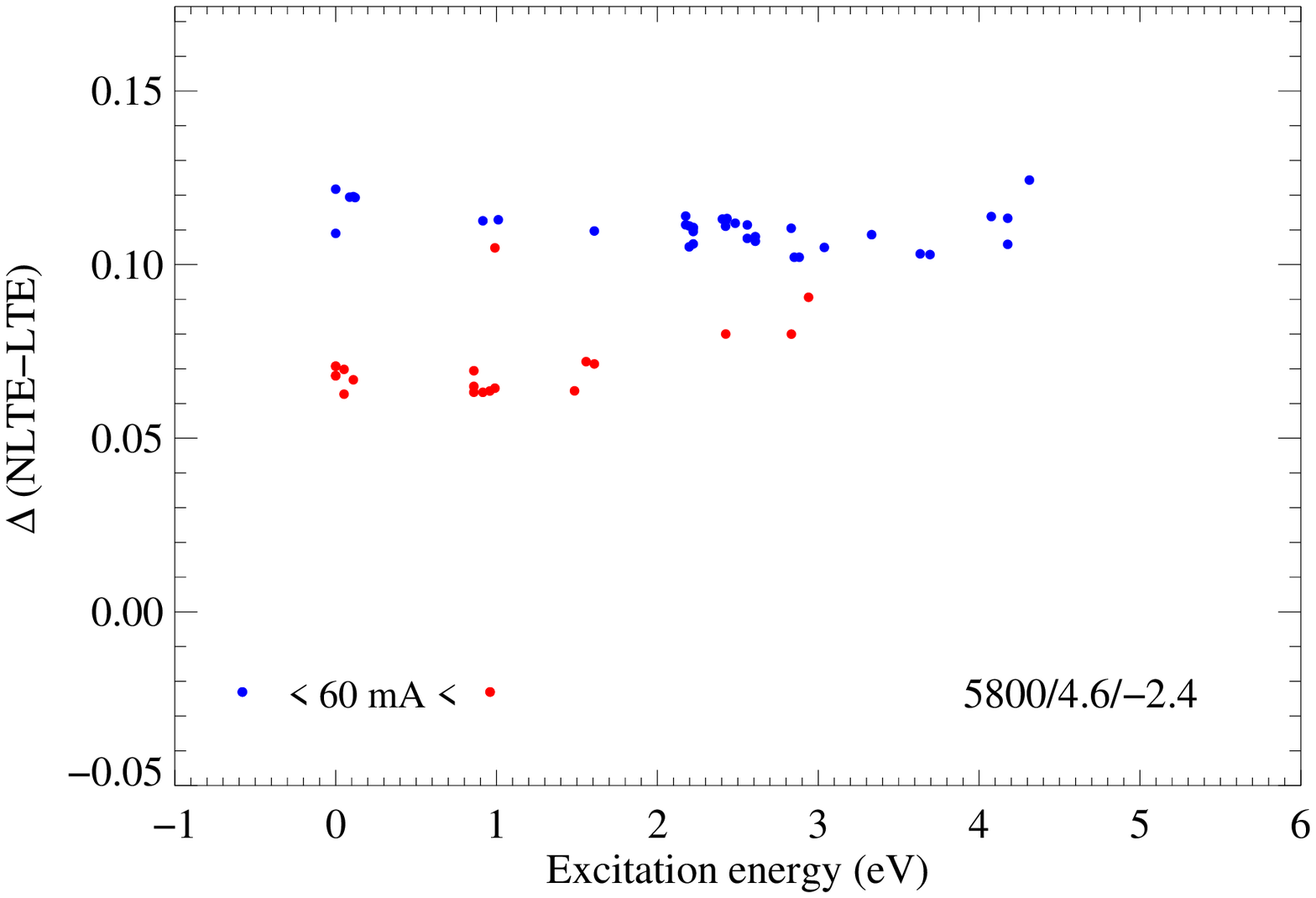}
}
\caption{\label{exc}NLTE abundance corrections for the Fe I lines as a function of lower level excitation potential. Red and blue colors separate the spectral lines with $\EW$ greater and less than $60$ \mA, respectively. Stellar parameters are indicated in the plots.}
\end{figure}

On the other side, non-negligible error bars indicate that there are line-to-line variations of NLTE abundance corrections for the same model atmosphere. The spread of $\Delta_{\rm NLTE}$ among Fe I lines with different equivalent widths $\EW$ and excitation potentials $\Elow$ strongly increases with decreasing metallicity for giants, and less so for dwarfs. Furthermore, this scatter is not random, but corresponds to a trend of $\Delta_{\rm NLTE}$ with $\Elow$, as shown in Fig. \ref{exc}. The figure demonstrates variation of NLTE corrections with lower level excitation potential for the models with parameters: $\Teff = 5000$, $\log g = 2.2$, [Fe/H] $= -2.4$ (left panel) and $\Teff = 5800$, $\log g = 4.6$, [Fe/H] $= -2.4$ (right panel). Stronger lines, $\EW > 60$ \mA, tend to experience smaller departures from LTE compared to weaker lines, but they also demonstrate systematically increasing $\Delta_{\rm NLTE}$ with increasing $\Elow$. No such effect is seen for the weak lines, $\EW \leq 60$ \mA, which, for some exceptions, have very similar NLTE corrections independent of their excitation potential.

Thus, our results suggest that LTE $\Teff$ determinations become very inaccurate at [Fe/H] $< -2$ and there are systematic effects in the $\Teff$ scale of dwarfs and giants.

\section{Conclusions}

We have constructed a new NLTE Fe model atom using the most up-to-date theoretical and experimental atomic data. The model has been tested on a number of well-studied late-type stars with parameters determined by other independent methods (Bergemann et al, in preparation).

Our preliminary results for the NLTE effects on Fe I and Fe II in the atmospheres of late-type stars are qualitatively consistent with other studies. Kinetic equilibrium of Fe, which is a minority ion at the typical conditions of these cool and dense atmospheres, favors lower number densities of Fe I compared to LTE. The number densities of Fe II are hardly affected by NLTE. In general, this leads to a weakening of Fe I lines compared to LTE, which, in turn, requires larger Fe abundance to fit a given observed spectral line. The magnitude of departures and NLTE corrections critically depends on stellar parameters. Whereas NLTE abundance corrections for the Fe I lines do not exceed $\sim 0.05$ dex for solar metallicity dwarfs, they can be as large as $0.6$ dex for metal-poor giants.

We find that both excitation and ionization balance is affected by NLTE effects on Fe I. NLTE corrections to Fe I lines increase with decreasing [Fe/H] and $\logg$, thus ionization balance achieved assuming LTE leads to progressively underestimated gravities and metallicities. On the other side, Fe I levels are also not in thermal equilibrium with respect to each other, and for the stronger lines NLTE abundance corrections show trends with line excitation potential. In particular, we find that LTE $\Teff$ determinations become very inaccurate at low metallicities and significant systematical effects in the $\Teff$ scale of dwarfs and giants can be expected.

The NLTE abundance corrections will be publicly available through an interactive online database, which is currently under construction. They can be used to retrieve NLTE abundances and abundance corrections for input equivalent widths. We also consider an option to provide functional dependency of NLTE corrections on line parameters for a grid of stellar parameters ($T_{\rm eff}$, $\log g$, [Fe/H]). This is a more convenient solution for implementation in the automated codes used in large-scale spectroscopic studies, such as follow-up spectroscopy for Gaia targets.

\ack
We would like to thank Luca Sbordone for the help with revision of the SIU code.

\section*{References}
\medskip
\begin{thereferences}

\item Allen C W 1973, {\it Astrophysical Quantities} {\bf 3rd ed} Athlone Press, London
\item Athay R G \& Lites B W 1972 {\it ApJ} {\bf 176} 809 
\item Bautista M A 1997 {\it A\&AS} {\bf 122} 167
\item B{\"o}hm-Vitense E 1958  {\it Zeitschrift f\"ur Astrophysik} {\bf 46} 108 
\item Butler K and Giddings J 1985 {\it Newsletter on Analysis of Astronomical Spectra} University of London {\bf 9}
\item Drawin H W 1969 {\it Z. Physik} {\bf 225} 470
\item Fuhrmann K, Pfeiffer M, Frank C, Reetz J, and Gehren T 1997 {\it A\&A} {\bf 323} 909
\item Grupp F 2004 {\it A\&A} {\bf 420} 289
\item Mashonkina L, Gehren T, Shi J R, Korn A J, and Grupp F 2011 {\it A\&A} {\bf 528} 87
\item Ramsbottom C A, Hudson C E, Norrington P H, and Scott, M P 2007 {\it A\&A} {\bf 475} 765
\item Reetz J K 1999 {\it PhD thesis, LMU Munich}
\item Steenbock W and Holweger H 1984 {\it A\&A} {\bf 130} 319
\item Takeda Y 1994 {\it PASJ} {\bf 46} 53
\item van Regemorter H 1962 {\it ApJ} {\bf 136} 906

\end{thereferences}

\end{document}